\documentclass[aps,groupedaddress,showpacs,letterpaper,twocolumn]{revtex4}

\usepackage{graphicx} 
\usepackage{amsmath} 
\usepackage{color}
\usepackage{ulem}
\begin{document}

\newcommand {\edit}[1]{\textcolor{red}{#1}}
\newcommand {\note}[1]{\textcolor{blue}{\textbf{#1}}}

\newcommand \be{\begin{equation}}
\newcommand \ee{\end{equation}}
\newcommand \bea{\begin{eqnarray}}
\newcommand \eea{\end{eqnarray}}
\newcommand \bse{\begin{subequations}}
\newcommand \ese{\end{subequations}}
\newcommand \bml{\begin{subequations}\begin{eqnarray}}
\newcommand \eml{\end{eqnarray}\end{subequations}}

\newcommand \mcA{{\mathcal A}}
\newcommand \mcE{{\mathcal E}}
\newcommand \nn{\nonumber}

\newcommand{\ket}[1]{| #1 \rangle}
\newcommand{\bra}[1]{\langle #1 |}
\newcommand{\braket}[1]{\langle #1 \rangle}

\title{Spin squeezing of atomic ensembles by multi-colour quantum non-demolition measurements}
\author{M. Saffman$^{1}$, D. Oblak$^2$, J. Appel$^2$, and E. S. Polzik$^2$}
\affiliation{
$^1$ Department of Physics,
University of Wisconsin, 1150 University Avenue,  Madison, Wisconsin 53706\\
$^2$ Danish National Research Foundation Centre for Quantum Optics (QUANTOP), Niels Bohr Institute, Copenhagen University, Blegdamsvej 17, DK- 2100 Copenhagen \O, Denmark}
 \date{\today}

\begin{abstract}
We analyze the creation of spin squeezed atomic ensembles by simultaneous dispersive interactions with several optical frequencies. 
A judicious choice of optical parameters enables 
optimization of an interferometric detection scheme  that  suppresses   inhomogeneous light shifts and keeps the interferometer operating in a balanced mode that minimizes technical noise. 
We show that when the atoms interact with two-frequency light tuned to cycling transitions the degree of spin squeezing $\xi^2$ scales as $\xi^2\sim 1/d$ where $d$ is the resonant optical depth of the ensemble.  In real alkali atoms there are loss channels and the scaling may be closer to  $\xi^2\sim 1/\sqrt d.$
Nevertheless  the use of two-frequencies provides a significant improvement in the degree of squeezing attainable as we show by quantitative analysis of non-resonant  probing on the Cs D1 line. 
Two alternative configurations are analyzed: a Mach-Zehnder interferometer that uses spatial interference, and an interaction with  multi-frequency amplitude modulated light that does not require a spatial interferometer.

\end{abstract}

\maketitle



\section{Introduction}

Coupling between light beams and atomic ensembles is of interest 
for processing and storing quantum information, and for enabling high precision measurements of fundamental physical quantities\cite{ref.polzikchapter}. 
Recent developments in  atomic clocks have demonstrated a measurement uncertainty that is limited by the quantum projection noise of atomic spin measurements\cite{salomonclock}. Reductions in the measurement uncertainty may be achieved by using spin squeezed states (SSS) of atomic ensembles\cite{uedasss,winelandsss}. Such states were generated using an off-resonant quantum nondemolition (QND) interaction with  a coherent light beam\cite{kuzmichqnd,ref.polzik2001}. Preparation of an  atomic sample in a SSS
reduces the variance of a projective measurement of the  spin  by a factor of $\xi^2=1/(1+\kappa^2)$ below that of an ensemble prepared in a coherent spin state (CSS). Here $\kappa$ is a constant proportional to the light matter interaction strength.  Development of techniques for generating strongly squeezed atomic samples  is therefore of great interest as a route to improving the precision of atomic clocks. 

Spin squeezed states can be generated via a QND interaction 
described by a Hamiltonian of the form ${\mathcal H}_{\rm QND}\sim \hat F_z \hat S_z$, with $\hat {\bf F}$ and $\hat {\bf S}$ referring to spin degrees of freedom of the atoms and the light respectively.
In this paper we will focus on the situation where the atomic pseudospin $\hat {\bf F} $ is defined in the  basis 
$|3\rangle=|f=3,m_f=0\rangle$ and $|4\rangle=|f=4,m_f=0\rangle$  corresponding to the clock transition between $f=3$ and $f=4$ ground state hyperfine levels in Cs.
The optical pseudospin operator $\hat {\bf S} $  may be defined in a basis of  polarization, spatial, or frequency modes of the light. One of the challenges encountered in preparation of spin squeezing is the fact that the above QND Hamiltonian is only an approximation that neglects additional aspects of the light-matter interaction which serve to reduce the usable amount of squeezing. 
 For example in the case of polarization dependent 
optical Faraday rotation there are  nonlinear terms in the atomic tensor polarizability which give a non QND like interaction\cite{ref.kupriyanov2005,jessenclock} leading to decoherence of atomic superposition states. We  will consider $\hat z$ polarized light beams as was used in our recent observation of Rabi oscillations on the Cs clock transition\cite{ref.polzik2008a}. The $\hat z$ polarized light couples to the atomic basis states without any nonlinear terms. Nevertheless  spatial inhomogeneity of the light-atom coupling strength  leads to inhomogeneous atomic phase shifts. It is still possible to obtain a  strongly squeezed ensemble 
characterized by a nonsymmetric entanglement measure in this situation\cite{kuzmichns}. However, the presence of inhomogeneous coupling is problematic in the context of  reducing projection noise in atomic clock experiments, since it limits the fidelity with which ensemble rotation operations can be 
performed\cite{ref.polzik2008b}. 

 In this paper we study the use of multi-frequency light beams for creating SSS. It was first shown in \cite{ref.mandel1998} that also with multiple probe frequencies a  QND interaction can be obtained. 
Here we consider a pair of two-frequency amplitude modulated light fields that are analogous to 
a carrier suppressed frequency modulation (FM) spectroscopy\cite{ref.bjorklundfm}. We use carrier frequencies symmetrically placed with respect to an atomic resonance to engineer an effective QND interaction,  while canceling inhomogeneous light shifts. In the spirit of 
Ref. \cite{kuzmichns} multiple frequencies encoded in a light beam with a common spatial mode provides a convenient method to ensure the matched interactions needed for nonsymmetric entanglement generation. 
We proceed in Sec. \ref{sec.interaction} by recalling the form of the interaction between an atomic ensemble and an off-resonant light field and estimate the degree of squeezing obtainable by probing of the Cs D1 line. 
We show that the interaction with a linearly polarized  single frequency probe beam suffers from inhomogeneous light shifts.
The inhomogeneous shifts can be eliminated using two probe beams of different frequencies in a Mach-Zehnder interferometer as discussed in Sec. \ref{sec.2probe}. Provided the number of photons is large compared to the number of atoms this interaction has the potential for producing SSS without unwanted inhomogeneous light shifts.  For an idealized light-atom interaction the two-frequency technique leads to 
 spin squeezing  that scales as $\xi^2\sim 1/d$ where $d$ is the resonant optical depth of the ensemble. Although loss channels on the Cs D1 line limit the asymptotic squeezing to $\xi^2\sim 1/\sqrt d$ we show nevertheless that the quantitative performance is better than for  one-frequency probing.

 In Sec. \ref{sec.fm} we present an alternative configuration which eliminates the spatial Mach-Zehnder interferometer in favor of frequency domain phase shifts. This   is attractive since it removes the requirement of mechanical stability inherent in using an interferometer.   We conclude in Sec. \ref{sec.conclusion} with a discussion of the results obtained. 


\section{QND measurement with a single probe beam}
\label{sec.interaction}

\begin{figure}[!t]
  \centering
  \includegraphics[width=8.5cm]{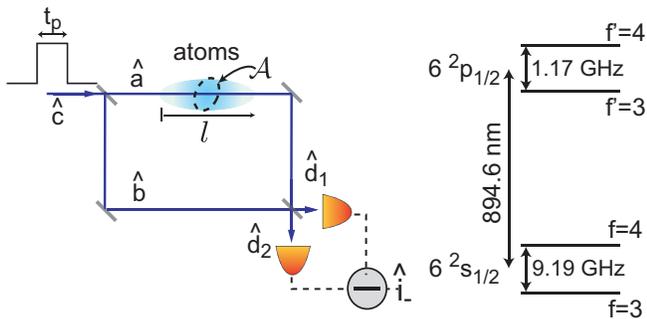}
\caption{(color online) Interferometric setup for QND measurement of atomic spin on the Cs D1 line. 
The coherent state input mode $\hat c$ is split into two equal intensity parts at the first beamsplitter, and then recombined at the output beamsplitter   to give output modes $\hat d_1,\, \hat d_2$ that are measured by photodetectors to generate a difference current $\hat i_-$.   }
  \label{fig.ae.mz}
\end{figure}

Consider an interferometric measurement of the collective atomic spin using the arrangement shown in Fig. \ref{fig.ae.mz}. This type of setup has been analyzed in detail in several papers\cite{moelmerib,oblak} and we will limit ourselves to a brief discussion in order to define notation to be used in what follows. 
 An input beam in a coherent state is split into two parts with equal amplitudes. A cloud of atoms  placed in one arm of the Mach-Zehnder imparts a phase shift on the light. The phase shift is transformed into an electronic  signal by measuring the difference of the photocurrents of the output detectors. Essentially this configuration was used in Ref. \cite{ref.polzik2008a} except that the probing light was tuned close to the Cs D2 line, instead of the D1 line considered here. 

The QND interaction generating spin squeezing in our system is governed by the Hamiltonian ${\mathcal H}_{\rm QND}\sim \hat F_z (\hat N_{\rm ph}/2+\hat S_z)$ where $\hat F_z=\sum_{i=1}^{N_{\rm at}}\hat f_{z}^{(i)},$ and $\hat S_z$ are operators for the $z$ component of  collective atomic\cite{ref.fleischhauercollective} and photonic spins respectively and $\hat N_{\rm ph}$ is the photon number operator (for convenience we set  $\hbar=1$).
The dimensionless single particle pseudospin operators  are
$
\hat f_x^{(i)}=\frac{1}{2}\left(\hat\rho_{34}^{(i)}+\hat\rho_{43}^{(i)} \right),$
$\hat f_y^{(i)}=-\frac{i}{2}\left(\hat\rho_{34}^{(i)}-\hat\rho_{43}^{(i)}\right),$ 
 $\hat f_z^{(i)}=\frac{1}{2}\left(\hat\rho_{44}^{(i)}-\hat\rho_{33}^{(i)} \right)$ for the atoms, while the continuous operators describing the light field are written as 
$
\hat S_x=\frac{1}{2}\left(\hat a^\dag\hat b+\hat b^{\dag}\hat a \right)t_{\rm p},$
$ \hat S_y=-\frac{i}{2}\left(\hat a^{\dag}\hat b-\hat b^{\dag}\hat a \right)t_{\rm p},$
$ \hat S_z=\frac{1}{2}\left(\hat a^{\dag}\hat a-\hat b^{\dag}\hat b \right)t_{\rm p}$.
Here  $\hat\rho_{jk}^{(i)}$ are ground state matrix elements of the single atom  slowly varying   density operator, $\hat a$ is the annihilation operator for the field that interacts with the atoms, 
and $\hat b$ is the annihilation operator for the local oscillator field in the lower arm of the interferometer which has no direct interaction with the atoms. 
The above definitions are supplemented by the  number operators for atoms and photons: 
$\hat N_{\rm at}= \sum_{i=1}^{N_{\rm at}}\left(\hat\rho_{33}^{(i)}+\hat\rho_{44}^{(i)}\right) $ and $\hat N_{\rm ph}= t_{\rm p} \left(\hat a^{\dag} \hat a+\hat b^{\dag} \hat b\right),$ with $t_{\rm p}$ the duration of the light pulse.

\begin{figure}[!t]
  \centering
  \includegraphics[width=8.cm]{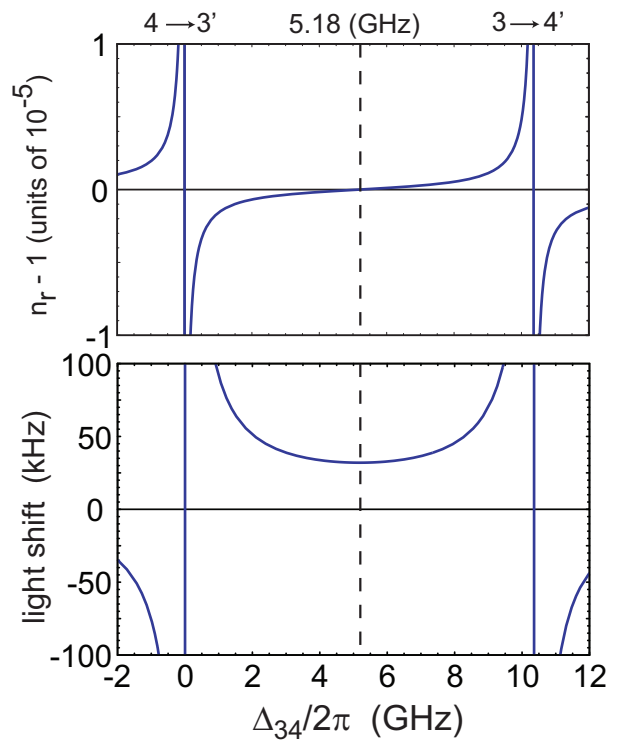}
\caption{(color online) Refractive index  and differential light shift near the D1 line for $\hat z$ polarized light,  an atomic  density of $1 \times 10^{11}~\rm cm^{-3}$ prepared with $\langle\hat F_z\rangle=0,$  $\mcA=\pi w^2$ with $w=20~\mu\rm m$, and an optical power of  $P=10^{-6}~\rm W.$ }
  \label{fig.shift}
\end{figure}

The probe beam refractive index and differential light shift of the clock states $|3\rangle, |4\rangle$ are shown in Fig. \ref{fig.shift} for light of wavelength $\lambda$ and frequency $\omega$ that is near resonant with the D1 line of Cs. 
A probe beam tuned to the zero index shift point indicated in Fig. \ref{fig.shift} receives equal and opposite phase shifts from population in the $f=3$ and $f=4$ states. The effective refractive index  is therefore proportional to the population difference of these states and is given by
$ n_r= 1-\frac{\lambda}{2\pi l}2\langle \hat F_z \rangle \tilde\kappa.
$
When the probe light is $\pi$ polarized  (linearly polarized along $\hat z$) the interaction constant is 
\bea
\tilde \kappa= \left(\frac{\lambda^2}{2\pi\mcA}\right)  
\frac{\frac{2\Delta_{34}}{\gamma}}{1+\frac{4\Delta_{34}^2}{\gamma^2}},
\eea
$\mcA$ is the  transverse area of the light beam and  atomic ensemble, $l$ is the length of the ensemble,   $\Delta_{jk}=\omega-(E_{6p_{1/2},f'=j}-E_{6s_{1/2},f=k})/\hbar$ is the 
detuning of the probe beam from the corresponding optical transition, and $\gamma$ is the radiative linewidth (FWHM) of the excited states. 
We assume that the atoms are  cold so that it is not necessary to account for the presence of Doppler broadening.

In the limit of large detuning where the photon scattering rate and accompanying population changes are small the  pseudospin operators transform as
$\hat {\bf F}^{\rm out}=\hat {\bf R}_z(\hat\theta_{\rm at})\hat {\bf F}^{\rm in},$
$\hat {\bf S}^{\rm out}=\hat {\bf R}_z(\hat\theta_{\rm ph})\hat {\bf S}^{\rm in},$ with
\bea
\hat {\bf R}_z(\hat\theta)
&=&
\begin{pmatrix}
\cos(\hat\theta)&\sin(\hat\theta )& 0\\
-\sin(\hat\theta)&\cos(\hat\theta )& 0\\
0&0& 1\\
\end{pmatrix}.
\nn
\eea
It follows from the form of  ${\mathcal H}_{\rm QND}$ that the  rotation angles are
\bse\bea 
\hat\theta_{\rm at}&=&2\tilde \kappa   (\hat N_{\rm ph}/2+\hat S_z^{\rm in} )\\
\hat\theta_{\rm ph}&=&-2\tilde \kappa  \hat F_z^{\rm in}.
\eea\label{eq.th1}\ese
These angles characterize the strength of the light-atom coupling.

\begin{figure}[!t]
  \centering
  \includegraphics[width=8.5cm]{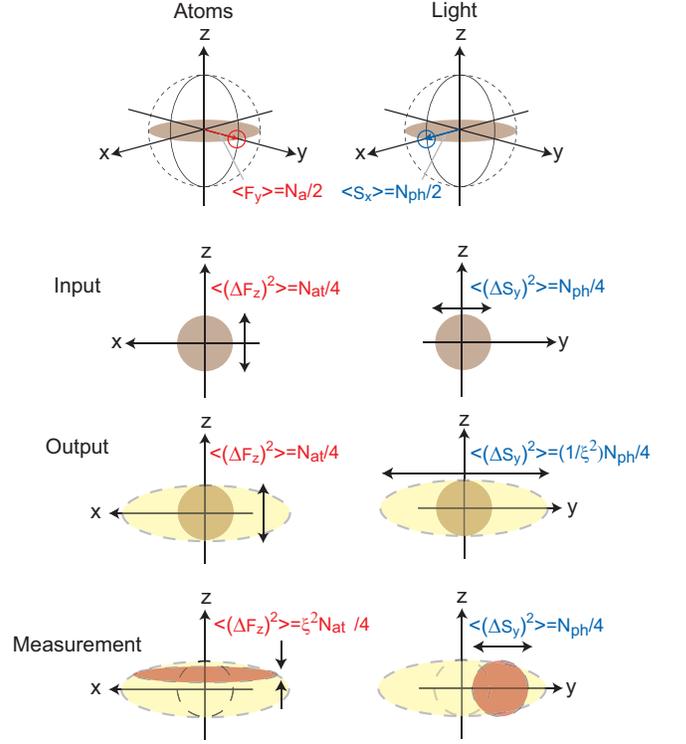}
\caption{(color online) Bloch sphere representation of CSS of atoms and light. The top two rows show the initial states and their fluctuations.
The output states after the interaction show that the atoms and light are rotated about the $z$ axis leading to unequal fluctuations transverse to the mean spin directions. Measurement of the light with quantum  limited uncertainty projects the atoms into a SSS with the variance of $\hat F_z$ reduced by a factor of $\xi^2.$ }
  \label{fig.ae.css}
\end{figure}

Consider  atoms and photons prepared in  CSS's as shown in Fig. \ref{fig.ae.css}. The atomic pseudospin is aligned such that  $\langle \hat F_x^{\rm in}\rangle = \langle \hat F_z^{\rm in}\rangle =0$, and $\langle \hat F_y^{\rm in}\rangle =\langle \hat N_{\rm at}\rangle/2=N_{\rm at}/2$. For the atoms, we may assume that we  prepared  the CSS  by starting with a  definite number of atoms in $|3\rangle$ so that initially $\langle\hat F_z\rangle=-N_{\rm at}/2$ and then used a perfect $\pi/2$ pulse to create the state with $\langle\hat F_y\rangle=N_{\rm at}/2.$   The  variances of the prepared state are  $\langle(\Delta \hat F_x^{\rm in})^2\rangle=\langle(\Delta \hat F_z^{\rm in})^2\rangle=N_{\rm at}/4$
and $\langle(\Delta \hat F_y^{\rm in})^2\rangle=0.$  
As for the light, the input port of the interferometer divides the  light equally between the two arms giving  
 $\langle \hat S_y^{\rm in}\rangle = \langle \hat S_z^{\rm in}\rangle =0$, $\langle \hat S_x^{\rm in}\rangle =\langle\hat N_{\rm ph}\rangle/2=N_{\rm ph}/2$  and input variances  $\langle(\Delta \hat S_x^{\rm in})^2\rangle=\langle(\Delta \hat S_y^{\rm in})^2\rangle=\langle(\Delta \hat S_z^{\rm in})^2\rangle=N_{\rm ph}/4.$ With these initial conditions we have $\langle \hat\theta_{\rm ph}\rangle=0$ but $\langle\hat\theta_{\rm at}\rangle \ne 0$ due to the presence of a nonzero differential light shift of the atomic states. We will return to the significance of the light shift below.

To lowest order in the interaction strength we find the output  variance of the light is
\bea
\langle(\Delta \hat S_y^{\rm out})^2\rangle&=&
\langle(\Delta \hat S_y^{\rm in})^2\rangle+  (2 \tilde \kappa )^2  \langle \Delta( \hat F_z^{\rm in }\hat S_x^{\rm in})^2\rangle\nn\\
&=&\frac{N_{\rm ph}}{4}\left(1+\kappa^2\right).
\label{eq.Syout}
\eea  
where
$\kappa^2 = \frac{1}{4} \tilde\kappa^2 N_{\rm at} N_{\rm ph}.$

To be of use, the operators for the light interacting with the atoms in one arm of a Mach - Zehnder interferometer as shown in Fig. \ref{fig.ae.mz}, must be related to a light observable at the interferometer output. To that end, we  choose the phase of the interferometer so that the powers in the two outputs are equal. At this position the interferometer has the greatest sensitivity to anticipated phase changes from the atomic sample and, additionally, any classical noise on the laser power is rejected. The field operators at the output beam splitter shown in Fig. \ref{fig.ae.mz} are related to the internal fields by 
$
\hat d_1=\frac{1}{\sqrt2}(\hat a+i\hat b), $
$\hat d_2=\frac{1}{\sqrt2}(\hat a-i\hat b).$
Consequently, the components of the light operator $\hat {\bf S}$ transform as 
$\hat S_{dx}=\hat S_z,$ 
$\hat S_{dy}=-\hat S_x,$ 
$\hat S_{dz}=-\hat S_y.$
where subscript $d$ refers to the interferometer output fields.  The output beams  are detected and the photocurrents subtracted. The difference photocurrent is proportional to $\hat \imath_-=\hat d_1^\dag \hat d_1 - \hat d_2^\dag \hat d_2=2 \hat S_{dz}=-2\hat S_y.$ {Combining with Eq.~(\ref{eq.Syout}) we have $\langle\hat \imath_- \rangle =0$
and 
\bea
\langle(\Delta \hat \imath_-)^2\rangle&=&4
\langle(\Delta \hat S_y^{\rm out})^2\rangle
=N_{\rm ph}(1+\kappa^2) .
\label{eq.dim}
\eea
When there are no atoms the variance of the measured difference current is given by $N_{\rm ph}$, the coherent state result as expected. 
When atoms are present $(\kappa^2>0)$ the variance  increases linearly with   the number of atoms, which is just the projection noise of a CSS.

A single quantum limited measurement of the difference photocurrent with variance $N_{\rm ph}$ represents a reduction by a factor of $1+\kappa^2$ compared to  the variance given by Eq. (\ref{eq.dim}). As is shown pictorially in Fig. \ref{fig.ae.css} the measurement projects the atoms into a spin squeezed state (SSS) with  the variance of the $z$ component reduced by the same factor, such that  
\bea
\langle(\Delta \hat F_z^{\rm out})^2\rangle&\rightarrow&
\frac{N_{\rm at}}{4}\frac{1}{ 1+   \kappa^2}.
\label{eq.dfzreduc}
\eea
The SSS is characterized by the degree of squeezing~\cite{uedasss,winelandsss} 
\be
\xi^2=\frac{\langle(\Delta \hat F_z^{\rm out})^2\rangle_{\rm SSS}}{\langle(\Delta \hat F_z^{\rm in})^2\rangle_{\rm CSS}}=\frac{1}{ 1+   \kappa^2}.
\label{eq.ae.xi}
\ee
A detailed discussion of the projective reduction of the atomic variance using a wavefunction formalism can be found in~\cite{moelmerib}. 
In order to  reduce the uncertainty in a Ramsey measurement of an atomic clock frequency additional operations are needed which include interchanging the variances of $\hat F_x$ and $\hat F_z$
as described in Ref. \cite{oblak}.

The degree of spin squeezing given by Eq. (\ref{eq.ae.xi}) neglects the deleterious effects of  inelastic light scattering which reduces the magnitude of the coherent spin state and adds noise to the $z$ components of the pseudo spins. It is well known\cite{ref.hammerer2004}
that the maximum attainable spin squeezing accounting for light scattering scales as $\xi^2\sim 1/\sqrt d$ with $d$ the resonant optical depth of the atomic sample. 
An exact calculation of the degree of spin squeezing in a real atomic system is very cumbersome. 
Previous work has provided  analytical results in a Gaussian approximation supplemented by numerical analysis to  account for atomic decay and light scattering\cite{ref.hammerer2004,ref.molmer2004}. The Gaussian state based calculations assume an idealized two-level atomic 
structure. The effect of atomic redistribution to other 
internal states due to light scattering was accounted for  in \cite{ref.polzik2005} 
for the case of $^{87}$Rb probed on the D2 line using an approximate analysis valid for not too large decay rates. Here we follow the spirit of \cite{ref.polzik2005} in the setting of the Cs D1 line. 

\begin{figure}[!t]
  \centering
  \includegraphics[width=8.5cm]{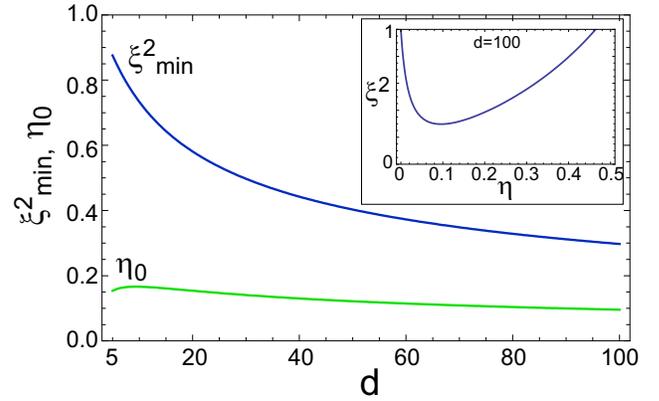}
\caption{Optimized spin squeezing and scattering parameter as a function of optical depth $d$
from Eq. (\ref{eq.ae.CsD1ss}). 
The inset shows the variation of $\xi^2$ with $\eta$ at $d=100.$ 
Realistic experimental parameters for Cs atoms in an optical dipole trap as in\cite{ref.polzik2008a} are $\lambda=0.894~\mu\rm m$,   $\mcA=\pi r^2$, $r=50~\mu\rm m$, $l=2 (\pi r^2/\lambda)=1.8~\rm cm$, atomic density $n_{\rm at}= 10^{10}~\rm cm^{-3}$. These values give $N_{\rm at}=1.4\times 10^6$ and $d=22$. With $N_{\rm ph}=9.5\times 10^{10}$ we get $\eta=0.15$ and $\xi^2=0.56.$ }
  \label{fig.ae.xisqmin}
\end{figure}

With  $\pi$ polarized light  tuned between the resonances as shown in Fig. \ref{fig.shift}
it is readily shown that $\kappa^2\simeq d\eta/2$ where $d=\sigma_0 N_{\rm at}/{\mathcal A}$, $\sigma_0=\lambda^2/2\pi$, and $\eta$ is the integrated probability that an atom scatters a photon during the probing pulse.  The result of the coherent QND  interaction can thus be written as $\xi^2=1/(1+\frac{1}{2}d\eta).$
Inelastic scattering events result in either decoherence and return of an atom to its original state with probability $\eta_{\rm dc}$ , or loss to states with $m_f=\pm 1$ with probability $\eta_l.$ These ``loss" states  couple to the probing light  with 
slightly different strengths. For the Cs D1 line and $\pi$ polarized light the coupling is about 6\% weaker for $m_f=\pm1$ than for $m_f=0$. As we have defined the atomic pseudospin in the basis of $m_f=0$ states we will consider population of $m_f\ne 0$ states as a loss mechanism and ignore the coupling of these states to the probing light.   In practice this assumption may be made realistic by  interspersing the QND interaction with cleaning steps that remove the population of $m_f\ne 0$ states. This could be done by, for example,
 coherently shelving the 
populations of the basis states to other levels and blowing away any population in the $m\ne 0$  states using resonant light. 
Accounting for the relevant  Clebsch-Gordan coefficients we find $\eta_l=2\eta/3, ~\eta_{\rm dc}=\eta/3$
and a short calculation then gives  
\be
{\xi}^2=\frac{1-\frac{2}{3}\eta}{1+\frac{1}{2}d\eta} +\frac{4}{3}\eta\frac{(1-\frac{2}{3}\eta)(1-\frac{3}{4}\eta)}{(1-\eta)^2}.
\label{eq.ae.CsD1ss}
\ee
For $d\gg 1$ the squeezing is optimized for $\eta_0\simeq\sqrt{3/2d}$ which gives 
${\xi}^2_{\rm min}\simeq\sqrt{32/3d}.$ Figure \ref{fig.ae.xisqmin} shows the degree of spin squeezing as a function of optical depth and scattering probability. We see that at $d=100$ the optimum is to set $\eta_0\simeq0.10$ which gives $\xi^2_{\rm min}\simeq0.30.$

The above discussion is still highly idealized in that it assumes a uniform interaction strength for all atoms in the sample. The zero phase shift frequency shown in Fig. \ref{fig.shift} imparts unequal light shifts to the clock states.  In a practical situation with a probing beam of Gaussian profile the strength of the light field will vary across the sample leading to inhomogeneous broadening and rapid loss of coherence between the clock states. The resulting dephasing of Rabi oscillations on the clock transition has been shown  to be well described by a model that accounts for a Gaussian beam profile and a Gaussian distribution of atoms in the ensemble\cite{ref.polzik2008b}.
In order to eliminate the inhomogeneous broadening it is necessary to use a probe frequency that results in equal light shifts for both clock states. This is possible using $\pi$ polarization on the D2 
line or linear polarization at an angle of $45^\circ$ from $\hat z$ on the D1 
line\cite{jessenclock}. Unfortunately, as shown in \cite{jessenclock}, the frequencies for which the light shifts are equalized are relatively close to resonance leading to strong photon scattering, and correspond to a nonzero phase shift of the light, so the interferometer operates in an undesired unbalanced configuration.


\section{QND measurement with two probe beams}
\label{sec.2probe}

We now show that it is possible to eliminate the inhomogeneous light shifts, and operate the interferometer  in a balanced configuration, while retaining the freedom of choosing the detuning to optimize the interaction strength.  To achieve this we use 
two $\pi$ polarized beams, one of  frequency $\omega_3$ tuned close to the $f=3\rightarrow f'={4}$ transition and one of frequency
$\omega_4$ tuned close to the $f=4\rightarrow f'={3}$ transition. 
 We introduce two sets of continuous operators $\hat S_{3x},\hat S_{3y},\hat S_{3z}$ and
$\hat S_{4x},\hat S_{4y},\hat S_{4z}$ for the two light fields respectively. These are defined in the same way as in the previous section with the replacements $\hat a\rightarrow \hat a_{3},$ 
$\hat b\rightarrow \hat b_{3},$ etc. . 

The light atom interaction is now characterized by four coupling constants corresponding to the interaction of each frequency with each of the ground states. We will be interested in detunings such that the 
interaction of $\omega_3$ light with population in $f=4$, and the interaction of $\omega_4$ light with population in $f=3$ is about one hundred times weaker than the interaction of each frequency with the population of the near resonant levels. We therefore only need to consider the two coupling constants to the near resonant levels
\bse\bea
\tilde \kappa_{3}&=&  \left(\frac{\lambda^2}{2\pi\mcA}\right)  
\frac{\frac{2\Delta_{43}}{\gamma}}{1+\frac{4\Delta_{43}^2}{\gamma^2}},\\  
\tilde \kappa_{4}&=&  \left(\frac{\lambda^2}{2\pi\mcA}\right)  
\frac{\frac{2\Delta_{34}}{\gamma}}{1+\frac{4\Delta_{34}^2}{\gamma^2}}, 
\eea\ese
where  $\Delta_{jk}=\omega_k-(E_{6p_{1/2},f'=j}-E_{6s_{1/2},f=k})/\hbar.$

\begin{figure}[!t]
  \centering
  \includegraphics[width=8.5cm]{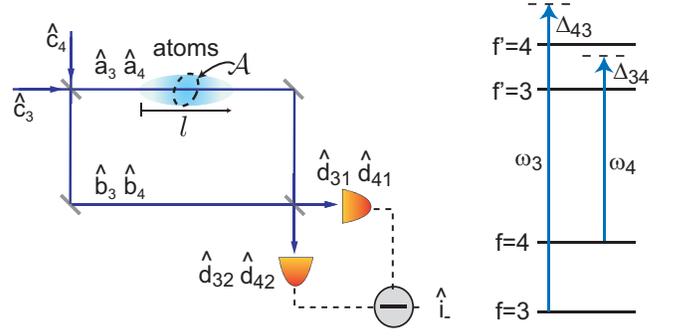}
\caption{(color online) Two frequency QND measurement of atomic spin. The probe frequencies are injected into different input ports as discussed in the text.   }
  \label{fig.twoportsin}
\end{figure}

Two conflicting requirements arise when using multiple probe beams. On the one hand we would like the total differential light shift to vanish. As seen in Fig.~\ref{fig.shift} this implies that $\Delta_{34}$ and $\Delta_{43}$ should have the same sign. 
On the other hand, in order to main to maintain equal intensities at both output ports of the interferometer,  which minimizes technical noise in the detection process, the two probe beams should acquire opposite phase shifts from their near resonant level implying that $\Delta_{34}$ and $\Delta_{43}$ should have opposite signs. We can solve this difficulty by choosing $\Delta_{34}, \Delta_{43}$ to have the same sign but let them be injected  into  different input ports as shown in Fig. \ref{fig.twoportsin}. Alternatively we could inject the two input fields $\hat c_3, \hat c_4$ into the same port, but with opposite circular polarizations and replace the input beamsplitter with a polarizing beamsplitter. A half-wave plate is then inserted into the arm that does not interact with the atoms to rotate by $\pi$ the polarizations of $\hat b_3$ and $\hat b_4$ so that they are aligned with the polarizations of $\hat a_3$ and $\hat a_4$ respectively when they are recombined at a nonpolarizing output beamsplitter. 
With either version the mean difference current is zero because the optical fields are phase shifted with opposite signs and their contributions to the difference current cancel.

We proceed by choosing $\Delta_{43}=\Delta_{34}$ which gives $\tilde\kappa_3=\tilde\kappa_4$.  Solving the Heisenberg equations for the light and atom evolution we find
$\hat {\bf F}^{\rm out}=\hat {\bf R}_z(\hat\theta_{\rm at})\hat {\bf F}^{\rm in},$
$\hat {\bf S_3}^{\rm out}=\hat {\bf R}_z(\hat\theta_{\rm ph,3})\hat {\bf S_3}^{\rm in},$
$\hat {\bf S_4}^{\rm out}=\hat {\bf R}_z(\hat\theta_{\rm ph,4})\hat {\bf S_4}^{\rm in},$
 with the  rotation angles 
\bse\bea
\hat\theta_{\rm at}&=&\tilde \kappa_4   [(\hat N_{\rm ph,4}/2+\hat S_{z4}^{\rm in} )-(\hat N_{\rm ph,3}/2+\hat S_{z3}^{\rm in} )]\\
\hat\theta_{\rm ph,3}&=&-\tilde \kappa_4 (\hat N_{\rm at}/2- \hat F_z^{\rm in})\\
\hat\theta_{\rm ph,4}&=&-\tilde \kappa_4 (\hat N_{\rm at}/2+ \hat F_z^{\rm in}).
\eea\label{eq.th2}\ese
The rotation angles are a factor of 2 smaller than those of Eqs. (\ref{eq.th1}) since now each frequency is assumed to only interact with one atomic ground state level.

The initial conditions for the mean values of the pseudospin operators are $\langle \hat F_y^{\rm in} \rangle=N_{\rm at}/2$,  $\langle \hat S_{3,x}^{\rm in} \rangle=N_{\rm ph,3}/2$,
 $\langle \hat S_{4,x}^{\rm in} \rangle=-N_{\rm ph,4}/2$, and all other components are zero. The opposite signs of the values of $\langle \hat S_{3,x}^{\rm in} \rangle$ and $\langle \hat S_{4,x}^{\rm in} \rangle$ is precisely a result of the injection of the two fields through two different input ports to the interferometer.
 Note that in contrast to the single probe beam situation we 
now have $\langle \hat\theta_{\rm at}\rangle=0 $ which implies that there is no differential Stark shift of the clock transition due to the probe beams, provided they have equal mean photon numbers which we will assume below. 
It follows immediately from Eq. (\ref{eq.Syout}) that the output variances of the light are
\be
\langle(\Delta \hat S_{3y}^{\rm out})^2\rangle=\langle(\Delta \hat S_{4y}^{\rm out})^2\rangle
=\frac{N_{\rm ph,4}}{4}\left(1+\kappa^2\right)
\label{eq.Syout2}
\ee 
where now $\kappa^2=\frac{1}{4}\tilde\kappa_4^2 N_{\rm at }N_{\rm ph,4}.$ 
The difference photocurrent at the output of the Mach-Zehnder is proportional to $\hat \imath_-=\hat d_{31}^\dag \hat d_{31} - \hat d_{32}^\dag \hat d_{32}+\hat d_{41}^\dag \hat d_{41} - \hat d_{42}^\dag \hat d_{42}=-2(\hat S_{3y}^{\rm out}+\hat S_{4y}^{\rm out}).$ The expected value of the difference current is  $\langle\hat \imath_- \rangle =0$ 
and the variance  is
\bea
\langle(\Delta \hat \imath_-)^2\rangle&=&4
\langle\Delta (\hat S_{3y}^{\rm out}+ \hat S_{4y}^{\rm out})^2\rangle\nn\\
&=& 2  N_{\rm ph,4}\left[ 1+2\kappa^2 \left(1+\frac{N_{\rm at}}{2 N_{\rm ph,4}}\right)\right].
\label{eq.2probevar}
 \eea
The variance includes a term proportional to $\kappa^2 N_{\rm at}/N_{\rm ph,4}$
which is  quadratic in the number of atoms.   The reason the variance was strictly linear in the number of atoms for the single probe beam, but has a quadratic correction for two probe beams, can be seen by comparing Eq. (\ref{eq.th1}b) with (\ref{eq.th2}b,\ref{eq.th2}c). 
In the two probe beam case the rotation angles include an additional factor of $\hat N_{\rm at}/2$. Each probe is phase shifted proportional to the (fixed) number of atoms. 
However, the noise of the photocurrent difference depends on 
 $\hat S_{3x}-\hat S_{4x}$  which has a  coherent state variance.  This gives a contribution to the measured photocurrent variance that is quadratic in the number of atoms.  Thus the two-probe technique is suitable for observing atomic projection noise in the limit when $N_{\rm at}/N_{\rm ph,4}\ll 1.$ Fortunately this limit is readily achievable in practice. 

From an experimental perspective the two input configuration has further advantages. Most importantly, the common mode noise e.g. arising from small displacements of the interferometer pathlength due to acoustics or vibrations, yields an opposite change in the output signal of the two probe fields. Hence, for equal power in the two input beams this noise will to first order be suppressed in the output photocurrent.
In the configuration where the probe fields enter the interferometer through two spatially separated input ports it is crucial that the fields have a very good spatial overlap in order that they interact with the atomic sample in exactly the same way. The configuration where the two fields enter the interferometer with orthogonal circular polarizations on a beamsplitter facilitates this mode overlap more readily as the fields may be spatially overlapped in a polarization-maintaining fiber before the interferometer.

In order to find the achievable spin squeezing we must again account for inelastic scattering. Before calculating the result for the Cs D1 line let us consider an idealized situation where 
the  probe at $\omega_3$ couples $\ket{3}\rightarrow \ket{3'}$ which only decays to $\ket{3}$ and the probe at 
 $\omega_4$ couples $\ket{4}\rightarrow \ket{4'}$ which only decays to $\ket{4}$. In this situation all moments of  the $\hat F_z$ operator are  unchanged by photon scattering and the reduction in spin squeezing is only due to a reduction in the magnitude of the coherent  spin state: $\braket{|\hat F_y|}\rightarrow (1-\eta)\braket{|\hat F_y|}.$ With the definition of spin squeezing relevant for Ramsey spectroscopy defined in \cite{winelandsss} we find 
\be
\xi^2=  \frac{1}{(1-\eta)^{2}}  \frac{1}{1+ d \eta}
\label{eq.sss2col}
\ee
where we have used $2\kappa^2=d\eta.$ 
From this equation one may find the inelastic scattering rate that yields the highest squeezing
\be
\eta_{0}=\frac{d - 2}{3d} , 
\label{eq.etaopt2col}
\ee
which for $d \gg 1$ gives $\eta_{0}\sim 1/3$. The corresponding maximal squeezing for large resonant optical densities is $\xi_{\rm min}^2=27/(4d).$ This $1/d$ scaling as opposed to the usual $1/\sqrt d$ is an attractive feature in the context of cold atomic samples with limited optical depth.

Such an  idealized situation is difficult to achieve in practice with available atomic level structures. One possibility is to use the basis states $\ket{F, m=\pm F}$ and the transitions $\ket{F, m=\pm F}\rightarrow \ket{F'=F+1,m'=\pm F\pm1}.$ Such states may not be well suited for atomic clocks since the clock frequency is defined by Zeeman shifts and will be both relatively small and linearly sensitive to magnetic field fluctuations. In the case of Cs  we may alternatively use two-colour probing of the clock states $\ket{3},\ket{4}$ with $\pi$ polarized light on  the D2 line. With frequencies chosen such that $f=3$ couples to $f'=2$ and $f=4$ couples to $f'=5$  the effect of photon scattering is to 
populate states with $m_f\ne 0$, but the value of $f$ is not changed. Since the states with $m_f\ne 0$ have a slightly different coupling strength to the light than the clock states some noise is added to the population difference measurement. It can be shown that for not too strong scattering such that we only need consider states with $m_f=\pm 1$ and optical depth not more than about $50$ the 
relation $\xi_{\rm min}^2=27/(4d)$ is still a good estimate to an accuracy of about 20\%. 
  Although we cannot achieve the asymptotic $1/d$ scaling of the spin squeezing for very large $d$, we nevertheless retain the other advantages of two-colour probing discussed above.

\begin{figure}[!t]
  \centering
  \includegraphics[width=8.5cm]{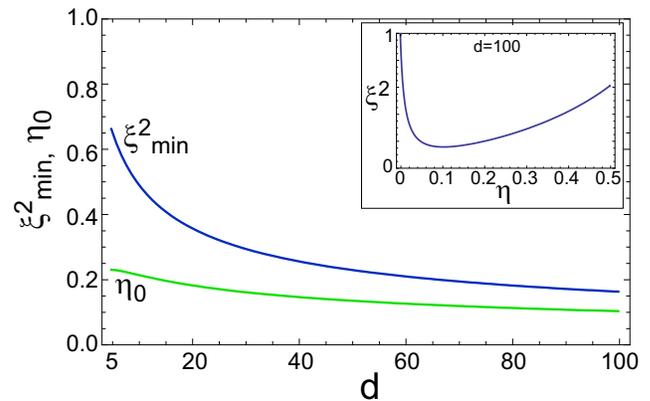}
\caption{Optimized spin squeezing and scattering parameter as a function of optical depth $d$
from Eq. (\ref{eq.ae.CsD1ss2}). 
The inset shows the variation of $\xi^2$ with $\eta$ at $d=100.$ 
With the same atomic parameters as in Fig. \ref{fig.ae.xisqmin} we have again $d=22$. Taking  $N_{\rm ph}=9.\times 10^{7}$ and $\Delta_{43}/2\pi=150~\rm MHz$ we get $\eta=0.17$ and $\xi^2=0.34.$ }
  \label{fig.ae.xisqmin2}
\end{figure}

For  a quantitative comparison with the one probe results of Fig. \ref{fig.ae.xisqmin}
let us again consider two-colour probing of the Cs D1 line as specified earlier in this section. 
Scattering of a photon at $\omega_3$ couples $\ket{3}$ to  $\ket{30}, \ket{31}, \ket{3-1},
\ket{41},$ or $\ket{4-1}$, with the kets labeled as $\ket{fm}$. Similarly scattering of  an $\omega_4$ photon couples $\ket{4}$ to 
$\ket{40}, \ket{41},$  $\ket{4-1}$, $ \ket{31},$ or  $\ket{3-1}.$ We denote the probabilities of these events by coefficients 
$\eta_{fm}^{(3)}$ for $\omega_3$ and $\eta_{fm}^{(4)}$ for $\omega_4$. 
For  the Cs D1 line we find  
$\eta_{30}^{(3)}=\eta/6, ~\eta_{3\pm1}^{(3)}=\eta/16, ~ \eta_{4\pm1}^{(3)}=5\eta/48,$ and
$\eta_{40}^{(4)}=\eta/6, ~\eta_{4\pm1}^{(4)}=5\eta/48, ~ \eta_{3\pm1}^{(4)}=\eta/16$.
The coefficients are normalized so that $\sum_{f,m} \eta_{fm}^{(3)}+\eta_{fm}^{(4)}=\eta.$ 
As in the discussion preceding Eq. (\ref{eq.ae.CsD1ss}) we assume that population in the $m\ne 0$ states is removed from the system. 
Calculating as in Refs. \cite{ref.molmer2004,ref.polzik2005}  we find  
\bea
\xi^2=\frac{(1-\frac{2}{3}\eta)^3}{(1-\eta)^2}\frac{1}{1+d\eta} +\frac{2}{3}\eta\frac{(1-\frac{2}{3}\eta)^2}{(1-\eta)^2}.
\label{eq.ae.CsD1ss2}
\eea
For $d\gg 1$ the optimum scattering probability scales as  $\eta_0\sim 1/\sqrt d$ and $\xi_{\rm min}^2\sim 1/\sqrt d.$ Figure \ref{fig.ae.xisqmin2} shows the calculated spin squeezing as a function of optical depth. 
We see that for large $d$ the spin variance is about twice smaller than for the single frequency probing 
of Fig. \ref{fig.ae.xisqmin}.


\section{Spin squeezing with amplitude modulated light}
\label{sec.fm}

\begin{figure}[!t]
  \centering
  \includegraphics[width=8.5cm]{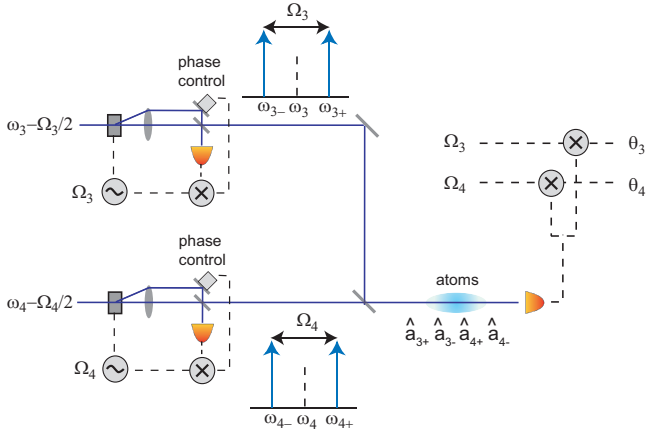}
\caption{(color online) Arrangement for spin squeezing with amplitude modulated light. Solid lines are optical beam paths and dashed lines are electrical signals. }
  \label{fig.ae.fm}
\end{figure}

An alternative approach to spin squeezing that does not rely on spatial interferometry is to 
measure the phase shifts between different frequency components of an amplitude modulated light field. 
The basic scheme is shown in Fig. \ref{fig.ae.fm}. By performing the phase sensitive measurements  in the frequency domain there is no need for the atoms to be placed inside a mechanical resonator that has  interferometric stability. This has the advantage that   
the relative phases of the different frequency components can be stabilized in a quiet environment that is physically separated from the atoms. 

We now have four  optical frequencies interacting with the atoms where the two $\omega_{3\pm}$ are symmetrically placed at a detuning of $\pm\Omega_{3}/2$ about $\omega_3$ and $\omega_{4\pm}$ are symmetrically detuned by $\pm\Omega_{4}/2$ about $\omega_4$. The central frequencies $\omega_3$ and $\omega_4$ are chosen such that the magnitude of the phase shift is equal for light at the lower and upper sideband, typically implying that $\omega_3$ and $\omega_4$ are set very near the resonance frequencies of the relevant atomic transitions. We emphasize that there is no light actually present at $\omega_3, \omega_4$
whereas  the 
interaction strength and photon scattering rate can be adjusted as desired by variation of the detunings
$\Omega_3, \Omega_4.$ 

The applicable continuous light operators are  combinations of the sideband frequencies interacting with each transition. We introduce photon annihilation operators 
$ \hat a_{3+}, \hat a_{3-}, \hat a_{4+}, \hat a_{4-}$ and construct the operators 
\bea
\hat S_{3x}&=&\frac{1}{2}\left(\hat a_{3+}^\dag\hat a_{3-}+\hat a_{3-}^\dag\hat a_{3+} \right)t_p\nn\\
\hat S_{3y}&=&-\frac{i}{2}\left(\hat a_{3+}^\dag\hat a_{3-}-\hat a_{3-}^\dag\hat a_{3+} \right)t_p\nn\\
\hat S_{3z}&=&\frac{1}{2}\left(\hat a_{3+}^\dag\hat a_{3+}-\hat a_{3-}^\dag\hat a_{3-} \right)t_p,\nn
\eea
and similarly with $3$ replaced by $4.$ The photon number operators once again are defined as $\hat N_{\rm ph,3}=\left(\hat a_{3+}^\dag\hat a_{3+}+\hat a_{3-}^\dag\hat a_{3-} \right)t_p,\nn$ and likewise for the $F=4$ ground state.

With the approximation of negligible photon scattering $\hat N_{\rm at}, \hat F_z,$ $\hat S_{3z}$, and $\hat S_{4z}$ are constants of the motion. 
We can therefore integrate the  Heisenberg equations for the light and atomic evolution to get
$\hat {\bf F}^{\rm out}=\hat {\bf R}_z(\hat\theta_{\rm at})\hat {\bf F}^{\rm in},$
$\hat {\bf S_3}^{\rm out}=\hat {\bf R}_z(\hat\theta_{\rm ph,3})\hat {\bf S_3}^{\rm in},$
$\hat {\bf S_4}^{\rm out}=\hat {\bf R}_z(\hat\theta_{\rm ph,4})\hat {\bf S_4}^{\rm in},$
 with the  rotation angles 
\bse\bea
\hat\theta_{\rm at}&=&\tilde \kappa_4  \hat S_{z4}^{\rm in} -
\tilde \kappa_3 \hat S_{z3}^{\rm in} \\
\hat\theta_{\rm ph,3}&=&\tilde \kappa_3 (\hat N_{\rm at}/2- \hat F_z^{\rm in})\\
\hat\theta_{\rm ph,4}&=&\tilde \kappa_4 (\hat N_{\rm at}/2+ \hat F_z^{\rm in}).
\eea\label{eq.thfm}\ese
where $\kappa_3$ and $\kappa_4$ are the common interaction strength of the respective pairs of sideband frequencies.
For equal powers in the sidebands the initial conditions for the light operators are $\langle \hat S_{3,x}^{\rm in} \rangle=N_{\rm ph,3}/2$,
 $\langle \hat S_{4,x}^{\rm in} \rangle=N_{\rm ph,4}/2$ with all other components equal to zero. As $\langle \hat S_{3,z}^{\rm in} \rangle=\langle \hat S_{4,z}^{\rm in} \rangle=0$ Eq. (\ref{eq.thfm}a) displays that there is no change in the expectation values of the components of atomic spin $\hat {\bf F}$ even when $\kappa_3 \neq \kappa_4$. In other words, the light shift is canceled by the combined influence of the two sidebands for each atomic level.

However, for the atomic output variance we find 
\bea
\langle(\Delta \hat F_x^{\rm out})^2\rangle&=&
\frac{N_{\rm at}}{4}\left(1+   \kappa^2\right),
\label{eq.ae.vfxofm}
\eea
where
\be
\kappa^2 = 4  \left(\tilde \kappa_3^2 N_{\rm ph,3} + \tilde\kappa_4^2 
N_{\rm ph,4}\right)   N_{\rm at} .
\label{eq.fmkappa}
\ee
The coupling constant $\kappa^2$ is effectively four times bigger than for the Mach Zehnder scheme. The reason being that now all the light interacts with the atoms. 

For the light, the photo detector measures the combined power of all the involved fields. The detector does not respond to the high frequency interference between $\omega_3$ and $\omega_4$ so the photocurrent operator is proportional to
\bea
\hat i&=&(\hat a_{3+}^\dag+\hat a_{3-}^\dag)(\hat a_{3+}+\hat a_{3-})+(\hat a_{4+}^\dag+\hat a_{4-}^\dag)(\hat a_{4+}+\hat a_{4-})\nn\\
&=&\hat N_{\rm ph,3} + \hat N_{\rm ph,4} +
\hat N_{\rm ph,3}\cos[\Omega_3 t + 4\tilde \kappa_3 (\hat N_{\rm at}/2-\hat F_z)]
\nn\\
&&
+ \hat N_{\rm ph,4}\cos[\Omega_4 t + 4\tilde \kappa_4 (\hat N_{\rm at}/2+\hat F_z)].
\eea

The photocurrent is split in two and mixed with the local oscillators $\Omega_3$ and $\Omega_4$ respectively to give two outputs 
\bea
 \hat N_{\rm ph,3}\sin[4\tilde \kappa_3 (\hat N_{\rm at}/2-\hat F_z)]
&\simeq& 
\hat N_{\rm ph,3} 4\tilde \kappa_3 (\hat N_{\rm at}/2-\hat F_z)=\theta_3\nn\\
 \hat N_{\rm ph,4}\sin[4\tilde \kappa_4 (\hat N_{\rm at}/2+\hat F_z)]
&\simeq& 
 \hat N_{\rm ph,4} 4\tilde \kappa_4 (\hat N_{\rm at}/2+\hat F_z)=\theta_4\nn
\eea
The difference of the measured phase angles is 
\bea
\theta&=&4 \hat N_{\rm ph,4}\tilde \kappa_4 (\hat N_{\rm at}/2+\hat F_z)
-4N_{\rm ph,3}\tilde \kappa_3 (\hat N_{\rm at}/2-\hat F_z)\nn\\
&=&2  (\tilde\kappa_4\hat N_{\rm ph,4}-\tilde\kappa_3\hat N_{\rm ph,3})\hat N_{\rm at}
+
4  (\tilde\kappa_4\hat N_{\rm ph,4}+\tilde\kappa_3\hat N_{\rm ph,3})\hat F_z.\nn
\\
\eea

We can choose the coupling constants and intensities such that 
$\tilde\kappa_4 N_{\rm ph,4}=\tilde\kappa_3 N_{\rm ph,3}$.
With this choice 
\bea
\hat \theta \hspace{-.1cm}&=&\hspace{-.1cm}
2\tilde\kappa_4 (\hat N_{\rm ph,4} - \frac{N_{\rm ph,4}}{N_{\rm ph,3}}\hat N_{\rm ph,3})\hat N_{\rm at}+
8 \tilde\kappa_4 \hat N_{\rm ph,4}\hat F_z
\eea 
which has a variance
\bea
(\Delta\hat\theta)^2&=&
 \kappa^2 N_{\rm at} + 16 (\tilde\kappa_4 )^2 N_{\rm ph,4}^2 N_{\rm at}\nn
\\
&=&
 2N_{\rm ph,4} \left[ \kappa^2 \left( 1+\frac{N_{\rm at}}{2 N_{\rm ph,4}}\right)\right]
\eea
where $\kappa^2$ is defined in Eq. (\ref{eq.fmkappa}).
We must add to this the variance due to the shot noise of the light when no atoms are present and 
choosing for simplicity $N_{\rm ph,3}=N_{\rm ph,4}$ (implying that $\Omega_3$ and $\Omega_4$ are chosen such that $\tilde\kappa_4=\tilde\kappa_3$) the shot noise becomes $N_{\rm ph,3}+N_{\rm ph,4}=2N_{\rm ph,4}$, yielding an output variance
\bea
(\Delta\hat\theta)^2= 
2 N_{\rm ph,4} \left[1+ \kappa^2 \left( 1 \hspace{-.1cm}+\hspace{-.1cm}\frac{N_{\rm at}}{2 N_{\rm ph,4}}\right)\right].
\eea
Apart from a different numerical factor  we find the same result as in Eq. (\ref{eq.2probevar}) for the two-probe interferometer. 
A quantum limited measurement of the phase angle $\theta$ will project the atoms into a SSS with reduced variance of $\hat F_z$ and as in the two-probe case of Sec. \ref{sec.2probe} there is an additional contribution to the variance which scales as $N_{\rm at}/N_{\rm ph}.$   The effects of photon scattering enter in the same way as in Sec. \ref{sec.2probe}. Thus this four-frequency approach has the potential for good spin squeezing performance. The most challenging technical requirement is the need for quantum limited 
phase measurements at a frequency $\Omega_4=\omega_{4+}-\omega_{4-}$ which must be at least a few times larger than the excited state radiative linewidth  in order to keep the  photon scattering rate  sufficiently low. 


\section{Conclusion}
\label{sec.conclusion}

We have analyzed the use of multiple probe frequencies for generation of 
 spin squeezed atomic ensembles. Robust preparation of atomic spin squeezing requires suppression of technical noise and inhomogeneous light shifts, together with the freedom to choose the optical detuning in order to 
optimize the photon scattering rate for a given atomic sample size. QND interactions with a single probe frequency do not generally allow all of these requirements to be simultaneously met. We have shown here, using the Cs D1 line as a specific example,  that the use of multiple frequencies with symmetrically chosen detunings can satisfy all of the above conditions simultaneously.
Two possible configurations were presented. The first uses a Mach-Zehnder interferometer as in recent non-destructive measurements of the Cs clock transition\cite{ref.polzik2008a,ref.polzik2008b}. The second configuration uses only frequency domain instead of spatial interference which  has the advantage of not requiring a mechanically stable interferometer.

We have also pointed out that with  multi-colour probing on  cycling transitions
the spin squeezing variance scales as $1/d$  as opposed to the $1/\sqrt{d}$ scaling obtained with single colour probing.  This type of multi-colour probing has been used in a recent demonstration of squeezing on the Cs clock transition\cite{Appel2008}.

\begin{acknowledgments}
The authors thank Anders S. S\o rensen for helpful discussions on the representation of decoherence.
This research was supported by the EU grants COMPAS and QAP. M. S. is grateful to the members of QUANTOP for hospitality during the preparation of this paper and acknowledges support from NSF and ARO-IARPA. 
\end{acknowledgments}


\end{document}